\documentclass[useAMS,usegraphicx,usenatbib]{mn2e}

\title [Metallicity effects on the cosmic SNIb/c and GRB rates] {Metallicity effects on the cosmic SNIb/c and GRB rates}

\author[V. Grieco et al.] {V. Grieco$^{1}$\thanks{E-mail:grieco@oats.inaf.it} F. Matteucci$^{1,2}$ G. Meynet$^{3}$ F. Longo$^{1}$ M. Della Valle$^{4}$  R. Salvaterra$^{5}$\\
$^{1}$Dipartimento di Fisica, Sezione di Astronomia, 
Universit\`a di Trieste, via G.B. Tiepolo 11, I-34131, Trieste, Italy \\
$^{2}$I.N.A.F. Osservatorio Astronomico di Trieste, via G.B. Tiepolo 11, I-34131, Trieste, Italy\\
$^{3}$Observatory of the University of Geneva, CH–1290 Versoix, Switzerland\\
$^{4}$I.N.A.F. Osservatorio Astronomico di Capodimonte, Salita Moiariello, 16, 801313, Napoli, Italy\\
$^{5}$I.N.A.F. IASF-Milano, via Bassini 15, I-20133, Milano, Italy}

\begin{document}
\date{Accepted . ; in original form 2011}

\pagerange{\pageref{firstpage}--\pageref{lastpage}} \pubyear{2002}

\maketitle

\label{firstpage}

\begin{abstract}
Supernovae Ib/c are likely to be associated to long GRBs, therefore it is important to compare the SN rate in galaxies with the GRB rate. 
To do that we computed Type Ib/c SN rates in galaxies of different morphological type (ellipticals, spirals and irregulars) by 
assuming different histories of star formation and different supernova Ib/c progenitors. We included some recent suggestions about 
the dependence of the minimum mass of single Wolf-Rayet (WR) stars upon the stellar metallicity  and therefore upon galactic chemical evolution. 
We adopted several cosmic star formation rates (i.e. relative to a comoving unitary volume of the Universe) as functions of cosmic time, 
either observationally or theoretically derived, including the one computed with our galaxy models. Then we  computed the cosmic Type Ib/c 
SN rates.
Our results show that the predicted Type Ib/c SN rates in spirals and irregulars can well reproduce the present 
time observed rates if both single WR stars and massive binary systems are taken into account as Type Ib/c SN progenitors. 
The metallicity effects on the minimum mass for single WR stars are evident mainly in the early phases of galaxy evolution and 
do not influence substantially the predicted local Type Ib/c rates.
We derived the following conclusions: i) the ratio cosmic GRB - Type Ib/c rate varies in the range $10^{-2}-10^{-4}$ 
in the whole redshift range, 
thus suggesting that only a small fraction of all the Type Ib/c SNe gives rise 
to GRBs. ii) The metallicity dependence of Type Ib/c SN progenitors produces lower cosmic SN Ib/c rates at early times, for any 
chosen cosmic star formation rate.
iii) Different theoretical cosmic star formation rates, computed under different scenarios of galaxy formation, produce SN Ib/c cosmic rates 
which differ mainly at very high redshift. However, it is difficult to draw firm conclusions on the high redshift trend because of the large 
uncertainties in the data. iv) GRBs can be important tracers of star formation at high redshift if their luminosity function does not vary 
with redshift and they can help in discriminating among different galaxy formation models. 
\end{abstract}

\begin{keywords}
supernovae -- gamma-ray bursts -- galaxy evolution.
\end{keywords}

\section{Introduction}
Gamma Ray Bursts (GRB) are sudden and powerful gamma-ray flashes,
occurring at a rate of $\sim$ 1 per day in the Universe.
The duration of GRBs at MeV energies ranges from $10^{-3}$ sec to about $10^{3}$ sec, with long bursts being characterized by a 
duration $> 2$ seconds.   
In the past years, it has been established that some long GRBs are
associated to supernovae  (SNe) originating from the death of massive stars.
The GRBs have been associated with powerful supernovae Ib/c having energies in excess of the majority of 
such SNe and for this reason 
they have been called \textquotedblleft hypernovae\textquotedblright (Iwamoto
et al. 1998; see also Paczy$\grave{\rm n}$ski 1998).
In particular, most of the evidence points towards Type Ic SNe (see Woosley \& Bloom 2006, Hjorth \& Bloom 2011).
The \textquotedblleft collapsar\textquotedblright model proposed to explain long GRBs takes into account this 
phenomenological aspect and proposes a  Wolf-Rayet progenitor which undergoes core collapse, thus
producing a rapidly rotating black hole surrounded by an 
accretion disk which injects energy into the system and 
thus acts as a \textquotedblleft central engine\textquotedblright (Woosley 1993, MacFayden \& Woosley 1999, 
Zhang, Woosley \& MacFayden 2003). However, the collapsar can originate also in massive stars in binary systems, as suggested by 
several authors (e.g. Baron, 1992; Kobulnicky \& Fryer, 2007; Yoon et al. 2010). 
Detailed galaxy evolution models are able to predict the temporal behaviour of SN rates in  galaxies of all morphological types. 
Therefore, a comparison 
between theoretical SN Ib/c rates and observed GRB rates seems appropriate at the present time. Bissaldi et al. (2007) already 
attempted such a comparison but the data relative to GRBs were much less than at the present time and limited to lower redshifts.
At the present time, GRBs have been observed up to $z \sim 8.2$ (Salvaterra et al. 2009; Tanvir et al. 2009).
In this paper we aim at studying the behaviour of the SN Ib/c rate in galaxies and as a function of redshift and compare it with the most recently 
derived cosmic GRB rate. 
In computing the Type Ib/c SN rate in galaxies we will adopt both single Wolf-Rayet and massive binaries as GRB progenitors and we will consider a 
dependence of the SN Ib/c progenitors on the initial stellar metallicity, not considered in any previous similar work (e.g. Bissaldi et al. 2007).
Our intention is to make predictions for the rates of Type Ib/c supernovae at various redshift in spiral and irregular
galaxies. At the moment, no observations can constraint such predictions, but in the future new powerful observational devices as the James Webb 
telescope or ELT  will provide extensive and constraining observational constraints on those rates. 
The comparisons with the present predictions will then allow to confirm or reject the present predictions and will bring new clues
on the nature of the Type Ib/c supernovae progenitors and on the star formation histories in spiral and irregular galaxies. 
From the comparison between observed GRB cosmic rate and predicted cosmic SN Ib/c rate, we aim first to check whether
the present ratio  of GRB to that of SN Ib/c can be well reproduced by our models, second to see how this ratio may change with 
redshift in the frame of our model.
In Section 2 we will describe the chemical evolution model adopted to compute the evolution of galaxies of different morphological type, 
as well as the computation of the SN Ib/c rate. In Section 3 the computed SN Ib/c rates for irregular  and spiral galaxies of 
different masses are presented and compared to the observed rates. In section 4 we assemble different cosmic histories 
of star formation, including the one computed by means of our galaxy models, to compute the cosmic Type I b/c SN rate, then we compare 
this cosmic SN rate with the cosmic GRB rate. 
Finally, in Section 5 a discussion and some conclusions are presented.

\section[]{The chemical evolution models}
In order to compute the Type Ib/c SN rate in galaxies we need to know the galaxy star formation history. Galaxies of different 
morphological type (ellipticals, spirals, irregulars) are characterized by different star formation histories (see Matteucci 2001). 
In particular, ellipticals should have suffered an intense and short star formation episode, whereas spirals and irregulars should have had 
milder 
star formation rates (SFR) and are still forming stars now. Irregular galaxies must have suffered the mildest SFR since they  contain more gas 
than galaxies of other types. 
 Here,  we focuse on irregular and spiral galaxies for the following reasons: i) because the SNe Ib/c are observed only in 
star forming galaxies and ii) because the observations on the hosts of GRBs have revealed that long GRBs are associated with faint, 
blue and often irregular galaxies (Conselice et al. 2005; Fruchter et al. 2006; Tanvir \& Levan 2007; Wainwright et al. 2007; Li 2008)
and tend to occur in galaxies with low metallicities (Fynbo \& al. 2003, 2006a; Prochaska et al. 2004; Soderberg et al. 2004; Gorosabel et al. 2005;  
Berger et al. 2006; Savaglio 2006; Stanek et al. 2006; Wolf \& Podsiadlowsky 2007; Modjaz et al. 2008; Savaglio et al. 2009).

However, we do not exclude that this could be a selection effect (see e.g. Mannucci et al. 2011; Campisi et al. 2011). In particular, 
in the paper of Mannucci et al. (2011), the authors suggest that probably the region high-$z$/high mass is populated by the dark GRBs. 
This idea is partially confirmed by the observations of some dark GRB host galaxies provided by Kruhler et al. (2011), where
the mass of the dark GRB hosts seem to be higher than the mass of the normal GRB hosts.

To do that we use a detailed self-consistent chemical evolution model reproducing the majority of the properties of irregular and spiral galaxies.
The irregular galaxies play an important role in the study of chemical 
evolution and star formation owing to their simpler structure 
and lack of evolution compared to  spiral galaxies. We also adopt a relatively simple model for spirals without taking into 
account gradients along the disk.

We assume that both irregular and spiral galaxies assemble all of their mass by means of a 
continuous infall of pristine gas. This is certainly true for spiral disks such as that of the Milky Way 
(see e.g. Chiappini et al. 1997; 2001, Boissier \& Prantzos 1999). 
The basic equations are:
\begin{equation}˙
\dot G_i = -\psi(t)X_i(t) + R_i(t) + (\dot G_i)_{inf} - \dot G_{iw}(t) 
\end{equation}
where $G_i(t) = M_{gas}(t)X_i(t)/M_{inf}$ is the gas mass in the
form of an element i normalized to the present time total luminous infall mass.
The quantity $X_i(t) = G_i(t)/G(t)$ represents the abundance by
mass of an element i and by definition the summation over all
the elements present in the gas mixture is equal to unity. The
quantity $G(t) = M_{gas}(t)/M_{inf}$ is the total fractional mass of
gas. The quantity, $R_i(t)$ represents the rate at which the element $i$ is restored into the ISM by the dying stars. 
Finally, $(\dot G_i)_{inf}$ and  $\dot G_{iw}(t)$  represent the infall and wind rate, respectively.

The SFR ($\psi(t)$), namely the amount of interstellar gas, expressed in 
solar masses, 
turning into stars per unit time, is assumed to be continuous and defined as a Schmidt (1959) law:

\begin{equation}
 \psi(t)=\nu G(t)
\end{equation}
where the quantity $\nu$ 
is the star formation efficiency (SFE),
namely the inverse of the typical time-scale for star formation, and
is expressed in $\mbox{Gyr}^{-1}$. The SFE for spirals is assumed to be higher than in irregulars.

We explored one initial mass functions (IMF), in particular the Salpeter (1955) one (x=1.35 in the mass range
0.1-100 $M_{\odot}$).

A main assumption of the model for irregulars is the existence of galactic winds triggered by SN explosions. 
In particular, it is assumed that a fraction ($\sim 30\%$) of the initial blast wave energy of SNe is transformed into thermal gas energy and   
the wind starts when the thermal energy of gas equates the binding energy of gas The wind can be \textquotedblleft normal\textquotedblright.  
namely each element is lost at the same rate, or \textquotedblleft differential\textquotedblright, in the sense that some elements (metals for example) 
are preferentially lost (see Bradamante et al. 1998).
The galactic wind is likely to occur in these systems because of their relatively low potential well. 
Moreover, galactic outflows are observed in irregular galaxies (see e.g. Martin, 1999; Martin et al. 2002). 
To do that we follow the method described in 
Bradamante et al. (1998) and Yin et al.(2010); in particular, the wind rate is assumed to be proportional 
to the SFR through a free parameter $\lambda_{i}$ larger than zero:
\begin{equation}
\dot G_{iw}(t)= \lambda_{i}  \psi(t)
\end{equation}
where $i$ represents a specific chemical element.
To have a preferential loss of metals, as indicated by dynamical models (e.g. Mc Low \& Ferrara, 1999), 
we use a differential wind  in which $\lambda_{H}=\lambda_{He}=0.3$ and  $\lambda_i \sim 0.9$ for the other elements.

In the case of spirals, the galactic wind is less likely to occur, owing 
the deep potential well in which the spiral disks lie. In fact, in spiral disks it is more likely 
to have galactic fountains rather than galactic winds (Spitoni et al. 2009).

Finally, the assumed rate of infall is the same for irregulars and spirals
and follows the law:

\begin{equation}
(\dot G_i){inf}= {a\,X_i e^{-t/ \tau} \over M_{inf}},
\end{equation}
where $a$ is a suitable constant, derived by integrating eq. (3)  over the galactic lifetime, $(X_i)_{inf}$ is the abundance  
by mass of the element $i$ in the infalling gas, assumed to be primordial and $\tau$ is the infall timescale. 
This timescale is  expressed in Gyrs and defined as the characteristic time at which half of the total mass of 
the system has assembled.  The values of $\tau$ are derived as the best ones in order to reproduce the majority 
of the observational constraints. This timescale is different for galaxies of different morphological type, 
being quite short in spheroids and increasing for spirals and irregulars.

In Table 1 we show the adopted model parameters for a typical spiral galaxy (Milky Way-like) and for a typical irregular galaxy.
   The infall mass $M_{inf}$ is the mass that eventually would be accreted if no galactic wind would occur. 
Model Irr is described by an infall mass of $5\cdot10^9 M_{\odot}$ which 
is one order of magnitude lower than the infall mass of Model Sp. The infall timescale for the spiral galaxy 
is assumed to be 6 Gyr, although in the Milky Way disk the timescale for disk formation should have been shorter in the internal 
than in the external regions (inside-out formation, Matteucci \& Fran\c cois 1989; Chiappini et al. 1997); here we want only to show 
some averaged properties and 6 Gyr is an average timescale between the internal timescales ($\sim 2$ Gyr) and the external ones ($\sim 10-12$ Gyr).
The timescale for the formation of irregulars is assumed to be smaller (4 Gyr).
Always in Table 1 we report the wind parameter for irregular galaxies  
($\lambda_i$, eq. 3) and the SFE for irregulars and spirals. 
By following the work of Calura et al. (2009) we assumed that the SFE increases with galactic mass.

\begin{table}
\begin{center}
\begin{tabular}{|c|c|c|}
 \hline
                              &    Model Irr                &      Model Sp        \\  
 \hline
$M_{inf}\, [M_{\odot}]$       &  $5\cdot10^9$               &   $5 \cdot 10^{10}$  \\
 \hline
$\tau \,[\mbox{Gyr}]$         &       $3$                   &        $6$           \\
 \hline
$\lambda_i$                   &  $\mbox{differential}$      &     $\mbox{no wind}$ \\
\hline
$SFE$ \, $[\mbox{Gyr}^{-1}]$  &      $0.05$                 &         $2$          \\
\hline
\end{tabular}
\end{center}
\caption{Parameter sets used for describing our models:$M_{inf}$ is the final total assembled mass if nothing is lost, $\tau$ is the infall time scale,
$\lambda_i$ is the wind parameter (see eq. 3) and $SFE$ is the star formation efficiency.}
\label{tabSFRparam}
\end{table}

Finally, in order to compute the cosmic star formation rate (CSFR) and Type Ib/c SN rate we have also considered a model for a typical 
elliptical of $10^{11}$ luminous mass. This model predicts a short and intense burst of star formation which stops before 1 Gyr, 
owing to the occurrence of strong galactic winds which devoid the galaxy of gas. The SFE adopted for this elliptical is $10\, \mbox{Gyr}^{-1}$ 
implying that this galaxy assembles more quickly than the late type ones. 
The assumed IMF is the Salpeter one, as assumed also for the other galaxy types.
It is worth noting that this model, such as the others for spirals and irregulars, well reproduce the local properties of ellipticals 
(Calura \& Matteucci 2004; Pipino \& Matteucci 2004).
The star formation history of this elliptical is shown in Figure \ref{figSFRmie} 
together with the SFRs of a spiral and an irregular galaxy. 

\subsubsection{Nucleosynthesis and stellar evolution prescriptions} 
Since we computed the chemical evolution of these galaxies, in particular the evolution of the O abundance, we adopted the yields 
from massive stars by Woosley \& Weaver 
(1995), the yields from low and intermediate mass stars by van den Hoek \& Groenewegen (1997) and the yields from Type Ia SNe 
by Iwamoto et al. (1999), their model W7.
Concerning the progenitors of Type Ib/c SNe it has been suggested that they could be single Wolf-Rayet (WR) stars , 
namely stars which have lost most of their H and He envelope and
with masses larger than $M_{WR}$, whose value depends on the initial stellar metallicity. 
In fact, the mass loss in massive stars ($M \ge 10M_{\odot}$) increases with the initial metallicity in a way 
that $M_{WR}$ decreases with increasing metallicity (see Tables 2,3).
In other words, with a high rate of mass loss, even stars of 20-25$M_{\odot}$ can become WRs.
However, the progenitors of Type Ib/c SNe could also be massive stars in binary systems in the mass range 12-20 $M_{\odot}$ 
(e.g. Baron, 1992; Bissaldi et al. 2007) or 14.8-45$M_{\odot}$ (Yoon et al. 2010).
Here we consider both progenitors (see Smartt 2009) following in part the work of Bissaldi et al. (2007), 
but adopting recent prescriptions for the dependence of $M_{WR}$  on metallicity and a mass range 14.8-45$M_{\odot}$ for the total mass of binary systems.
In particular, we consider the results of Georgy et al. (2009) which 
give the variation of  $M_{WR}$  as a function of the metallicity for all the core collapse SNe.
In Tables 2 and 3 we show the  $M_{WR}$ - Z relations extrapolated from the results of 
Georgy et al. (2009) and adapted to the  metallicity range of our galactic models for SNIb/c  and only Ic progenitors, respectively. 
In particular, in column 1 we show the relation  between the minimum WR mass and metallicity for a range of Z values, indicated in the second column. 
Finally, in the third column we show the minimum WR mass corresponding to specific metallicities.

\begin{table}
\tiny
\begin{center}
\begin{tabular}{|l|c|r|}
 \hline
 $M_{WR}-Z$  relation               &    Z range                  &     $M_{WR}(M_{\odot})$\\  
 \hline 
 $M_{WR} = -15290 Z + 113.76 $      &   $Z \leq 0.004$            &  $\sim52.6M_{\odot} \: @ \; Z=0.004  $ \\
 \hline
 $M_{WR} = -5650 Z + 75.20 $        &   $0.004 < Z \leq 0.008$    &  $\sim30M_{\odot} \: @ \; Z=0.008  $ \\
 \hline
 $M_{WR} = -416.5 Z + 33.33 $       &   $0.008 < Z < 0.02 $       &  $\sim27.7M_{\odot} \: @ \; Z=0.0134 $ \\
 \hline
$M_{WR} = -230 Z + 29.6 $           &   $0.02 \leq Z \leq 0.04 $  &  $\sim25M_{\odot}   \: @ \; Z=0.020  $ \\
 \hline
$M_{WR} = 20 M_{\odot} $            &   $ Z > 0.04 $              &  $\sim20.4M_{\odot} \: @ \; Z=0.040  $ \\
 \hline
\end{tabular}
\end{center}
\caption{Relations between the minimal WR mass able to form SNeIb/c and metallicity, in different ranges of metallicity.}
\label{tabMZ}
\end{table}
\normalsize

\begin{table}
\tiny
\begin{center}
\begin{tabular}{|l|c|r|}
 \hline
 $M_{WR}-Z$  relation          &    Z range                  &     $M_{WR}(M_{\odot})$\\  
 \hline 
 $M_{WR} = -10759 Z + 116 $    &   $Z \leq 0.008$            &  $\sim30M_{\odot} \: @ \; Z=0.008  $ \\
 \hline
 $M_{WR} = 750 Z + 24 $        &   $0.008 < Z < 0.02$        &  $\sim34M_{\odot} \: @ \; Z=0.0134 $ \\
 \hline
 $M_{WR} = -700 Z + 53 $       &   $0.02 \leq Z \leq 0.04 $  &  $\sim39M_{\odot} \: @ \; Z=0.02  $ \\
 \hline
$M_{WR} = 25 M_{\odot} $       &   $ Z > 0.04 $              &  $\sim25M_{\odot} \: @ \; Z=0.04  $ \\
 \hline
\end{tabular}
\end{center}
\caption{Relations between the minimal WR mass able to form SNeIc and metallicity, in different ranges of metallicity.}
\label{tabMZIc}
\end{table}
\normalsize

\subsection{The computation of the supernova Ib/c rate} \label{SNR}

The distinguished features of Type Ib and Ic SNe is the lack of conspicuous hydrogen spectral lines. 
The SNe Ib/c occur preferentially
in the vicinity of star forming regions and their progenitors are thought to be massive stars
that have lost most of their H-rich (and perhaps their He-rich) envelopes via strong winds or transfer to a binary companion via Roche 
overflow.
Approximately 25\% of all Core Collapse SNe fall in the SNe Ib and SNe Ic category (Hamuy 2003).
 
The SN Ib/c  and SNIc rates have been calculated assuming both single WRs and  stars in close binary systems as progenitors. 
In general: 

\begin{eqnarray}
 SNR  = \int_{M_{WR}}^{100} \psi(t-\tau_M) \phi(M)dM + \nonumber
\end{eqnarray}

\begin{eqnarray}
+ F \int_{14.8}^{45} \psi(t-\tau_M) \phi(M)dM \nonumber
\end{eqnarray}
\smallskip
\begin{equation}
 \sim \psi(t) ( \int_{M_{WR}}^{100} \phi(M)dM + F \int_{14.8}^{45} \phi(M)dM )
\end{equation}
where the lifetime $\tau_M$  of massive stars is considered negligible, $\psi(t)$ is the star formation rate and  $\phi(M)$ is  
the IMF. 
The same equation is used for both SNe Ib/c and SNe Ic alone, and the only difference is in
the minimum stellar mass of the progenitors, $M_{WR}$, varying as shown in Table \ref{tabMZ} 
for the SNe Ib/c  and in Table 3 for the sole SNe Ic. 
Concerning  Type Ib/c SNe, in one case the evolution of  $M_{WR}$, shown in Figure  \ref{EvoZMwr}, 
is taken into account,
while in the other case  $M_{WR}$ is assumed to be independent from metallicity and to be 25$M_{\odot}$.
Since, in this last case, the evolution of the SNR is quite the same for both the SN Types,  
we do not  show the evolution of SNe Ic with $M_{WR}$ constant.
The factor F represents the fraction of massive binary stars producing Type Ib/c SNe. For the moment, this parameter is chosen to 
be equal to 0.15, (Calura \& Matteucci, 2006). 
This value is motivated by the facts that first, in any galaxy, half of the massive stars are possibly in binary systems, 
and second, the fraction of massive stars in close binary system is $\sim 30\%$, i.e. similar to the close binary frequency 
predicted for low mass systems (Jeffries \&  Maxted, 2005). 
Therefore, the estimated value for this parameter is given by:
$$F \sim 0.5 \cdot 0.3 \sim 0.15. $$
 This is in good agreement with Podsiadlowski et al. (1992) who calculated that 15-30\% of all massive
stars (with initial masses above $8M_{\odot}$) could conceivably transfer mass to an interacting
companion and end up as helium stars.
However, we have tested also other values of the F parameter: in particular, for spirals we run models with F in the 
range 0.01-0.5 and found that the error on the predicted present time SN Ib/c rate is $\sim 0.003$ SNe $yr^{-1}$, 
while for irregulars we run models with F in the range 0.01-0.3 (the value of 0.5 gives too high present time SN Ib/c 
rates relative to observations) with an error on the theoretical SN Ib/c rate of $\sim 0.0003$ SNe $yr^{-1}$. 
We can safely conclude that in both cases F values lower than 0.15 are not enough to reproduce the 
observed Type Ib/c SN rates (see next paragraph), 
whereas higher  but not unreasonable values (up to F=0.5), do not produce 
sensible differences in the results. Therefore, we do not exclude that our 
chosen value of F could be a lower limit.

\section[]{Results}

Before comparing model results with the observed properties of galaxies, 
here we summarize some observational facts which are 
used to constrain our models:
 \begin{enumerate}[i)]

\item the total fractional mass of gas in irregulars is:  
          $$\mathrm{(M_{gas}/M_{tot})_{t\sim13\, Gyr} = [0.2 - 0.8]}$$
whereas in spirals is:
         $$\mathrm{(M_{gas}/M_{tot})_{t\sim13\, Gyr} < 0.3}$$
    
\item the global average metallicity in irregulars is:
          $$\mathrm{Z(t\sim13\, Gyr) = [0.03 - 0.5]\,Z_{\odot}} $$
whereas the average metallicity in the disk of a Milky Way like spiral is:
          $$\mathrm{Z(t\sim13\, Gyr) =[2-2.5]\,Z_{\odot}} ,$$
where $Z_{\odot}\sim 0.0134$ (Asplund et al. 2009).
    
\item The stellar mass vs. metallicity relation  is an important observational constraint both for spirals and irregulars. 
In particular, these latter seem to be the lower part of the relation for spirals. 
The mass-metallicity relation indicates that the stellar mass of 
star forming galaxies is correlated with the galaxy metallicity: galaxies with larger stellar masses tend to have higher metallicities 
(Tremonti et al. 2004, Savaglio et al. 2005; Erb et al. 2006; Kewley \& Ellison 2008: Maiolino et al. 2008).
\item The predicted SN Ib/c rate (SNR Ib/c) should reproduce the following 
                observational rates provided by Li et al. (2011):
                $$SNR_{Ib/c}=0.103_{-0.067}^{+0.136} SNuM \; \mbox{(Irregulars)}$$
                $$SNR_{Ib/c}=0.113_{-0.025}^{+0.031} SNuM \; \mbox{(Spirals)}$$
                where $SNuM=SNe (100\, \mbox{yr})^{-1} (10^{10} \, M_{\odot})^{-1} $ 
is the SN Ib/c rate per unit mass, in good agreement with previous works (i.e. Mannucci et al. (2005)). 
These observed rates can therefore be computed for galaxies of the same stellar mass as in our 
models and compared with our predicted rates.
 \end{enumerate}

The predicted values at present time of the total metallicity (Z), total fractional mass of gas in the galaxy $(M_{gas}/M_{tot})$,
stellar mass in solar units $(M_{star})$  and oxygen abundance (expressed as $log(O/H)+12$) are showed in the table \ref{tabSFRresults}, 
respectively.
It should be noted that the values of Z and 12+ log(O/H) for the spiral galaxy are larger than solar, 
the reason is that they represent the average metallicities over the entire galactic disk, 
where the inner regions have oversolar values and the external ones have lower values.

We see that present models provides values in agreement with observations for
the present total mass fraction of gas and for the average metallicity in both types of galaxies considered here.

\begin{table}
\begin{center}
\begin{tabular}{|c|c|c|}
 \hline
                                             &    Model Irr      &      Model Sp    \\  
 \hline
Z                                            &     $0.0096$      &       $0.03$    \\
 \hline
$\mathrm{(M_{gas}/M_{tot})_{t\sim13\, Gyr}}$   &     $0.66$       &       $0.017$     \\  
 \hline
$M_{star} \:[M_{\odot}]$                     & $1.4\cdot10^9$   & $4.27\cdot10^{10}$ \\
 \hline
$12+log(O/H)$                                &     $8.6$         &       $9.12$     \\
 \hline
$SFR \:[M_{\odot}\, \mbox{yr}^{-1}]$         &     $0.16$        &       $1.67 $    \\
 \hline
\end{tabular}
\end{center}
\caption{Predictions of the Model Irr and Model Sp at present time: the total metallicity, 
         total fractional mass of gas in the galaxy, stellar mass, oxygen abundance and star formation rate.}
\label{tabSFRresults}
\end{table}

\begin{figure}
\begin{center}
\includegraphics[width=0.5\textwidth]{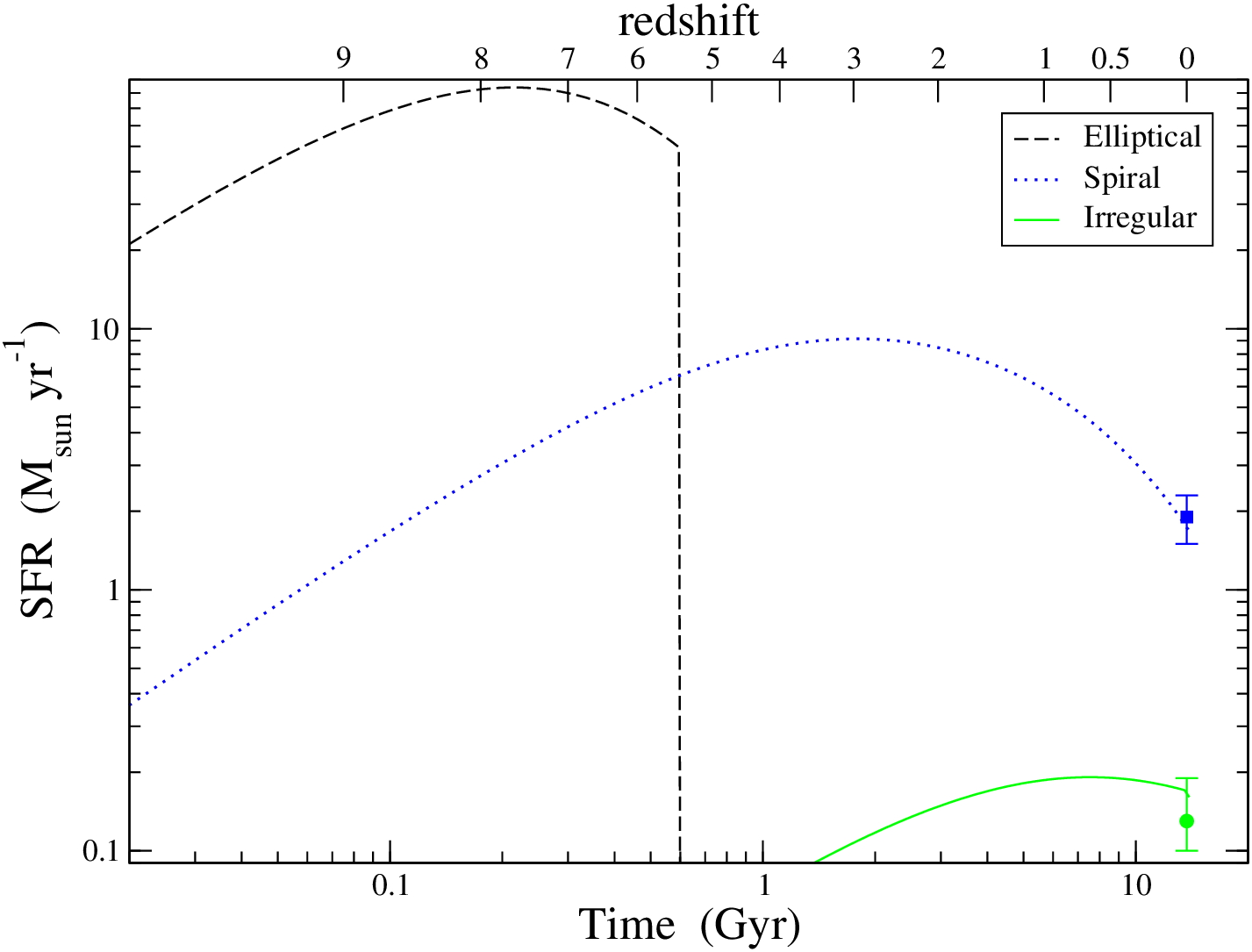} 
\caption {Predicted star formation rates of a typical elliptical, spiral and  irregular galaxy, 
expressed in $M_{\odot}/\mbox{yr} $  as  functions of time and redshift; 
the redshift of galaxy formation is $z_{f}=10 $ in a $\Lambda$CDM cosmology. 
The infall masses for each type of galaxies are: $10^{11}M_{\odot}$ 
(elliptical, dashed line) $5\cdot10^{10}M_{\odot}$ (Spiral, dotted line) and $5\cdot10^9M_{\odot}$ (Irregular,  solid line). 
 In the figure are shown also some average values for the present time SFR in spirals (square, Chomiuk \& Povich, 2011) 
and in irregulars (circle, Harris \& Zaritsky, 2009).}
    \label{figSFRmie} 
    \end{center}
    \end{figure}

In Figure \ref{figSFRmie}, the time evolution of the star formation rates (expressed in $M_{\odot}/\mbox{yr}$ ) is plotted for the 
three different types of galaxies (Elliptical, Spiral and Irregular).
The SFR increases until the energy injected into the ISM by stellar winds and SN(Ia, Ib, and II) explosions
triggers a galactic wind. At that time, the thermal energy is equivalent to the binding energy of gas, and the gas is 
lost at a rate proportional to the SFR (eq.3) with a consequent drop of the SFR.
  Moreover, we can see that the present models well fit the present time averaged SFR both in spirals and in irregulars.

\begin{figure}
    \begin{center}
    \includegraphics[width=0.5\textwidth]{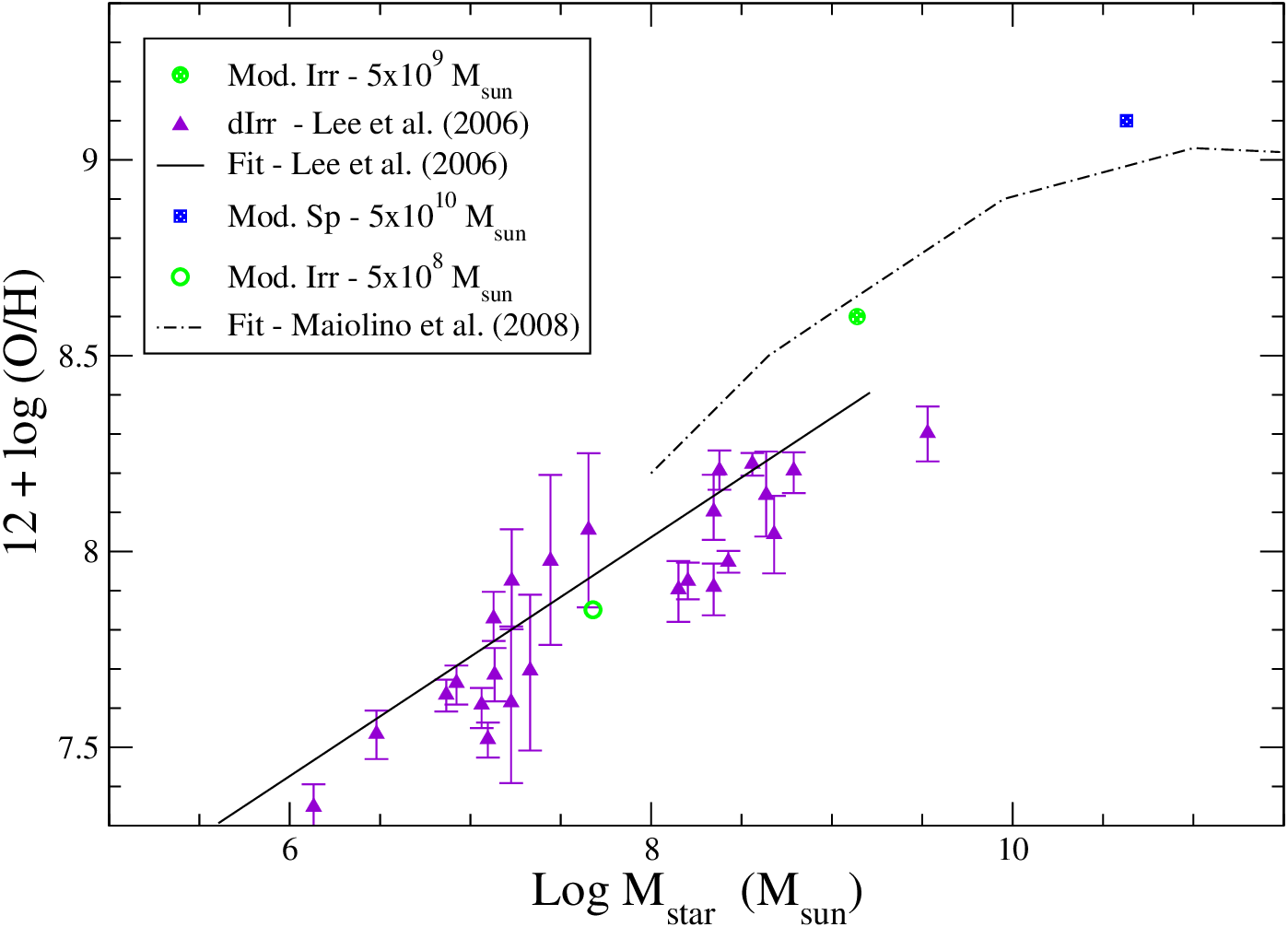} 
   \caption{Predicted and observed mass-metallicity relation for irregular 
galaxies; the continuous curve represents the best fit to
 the data of Maiolino et al. (2008) 
relative to star forming galaxies together with data and fit obtained by
 Lee et al. (2006) for a sample of local dwarf galaxies (purple triangles). The points are the predictions for Model Irr (green circles) and 
Model Sp (blue square). Note that for the irregulars we are showing also a model with infall mass $M_{inf}=5 \cdot 10^{8}$ and SFE=0.02 (open green circle).} 
    \label{figMZrel} 
    \end{center}
    \end{figure}

Figure \ref{figMZrel} shows the predicted and observed  mass-metallicity relation at the present time relative to small mass galaxies. 
In particular, we show both the best fit of Maiolino et al. (2008) of the data provided by Kewley \& Ellison (2008) 
concerning star forming galaxies  and the data and best fit of dwarf irregulars, as inferred by Lee et al. (2006).
As one can see, our  models  lie  close to the best fits. 
In our galaxy models the mass-metallicity relation arises naturally by adopting a smaller SFE in smaller galaxies. 
 The above comparisons show that our chemical models very well reproduce many observed properties of the present day spirals and
irregulars. Since, these present day properties results from the whole previous evolution, these good correspondence give 
some confidence that these models can be used to explore the much less well known early phases of these galaxies, 
which will hopefully be observed in greater details in a near future thanks to more powerful observational devices 
as the James Webb telescope and the ELT.

\begin{figure}  
    \begin{center}
    \includegraphics[width=0.5\textwidth]{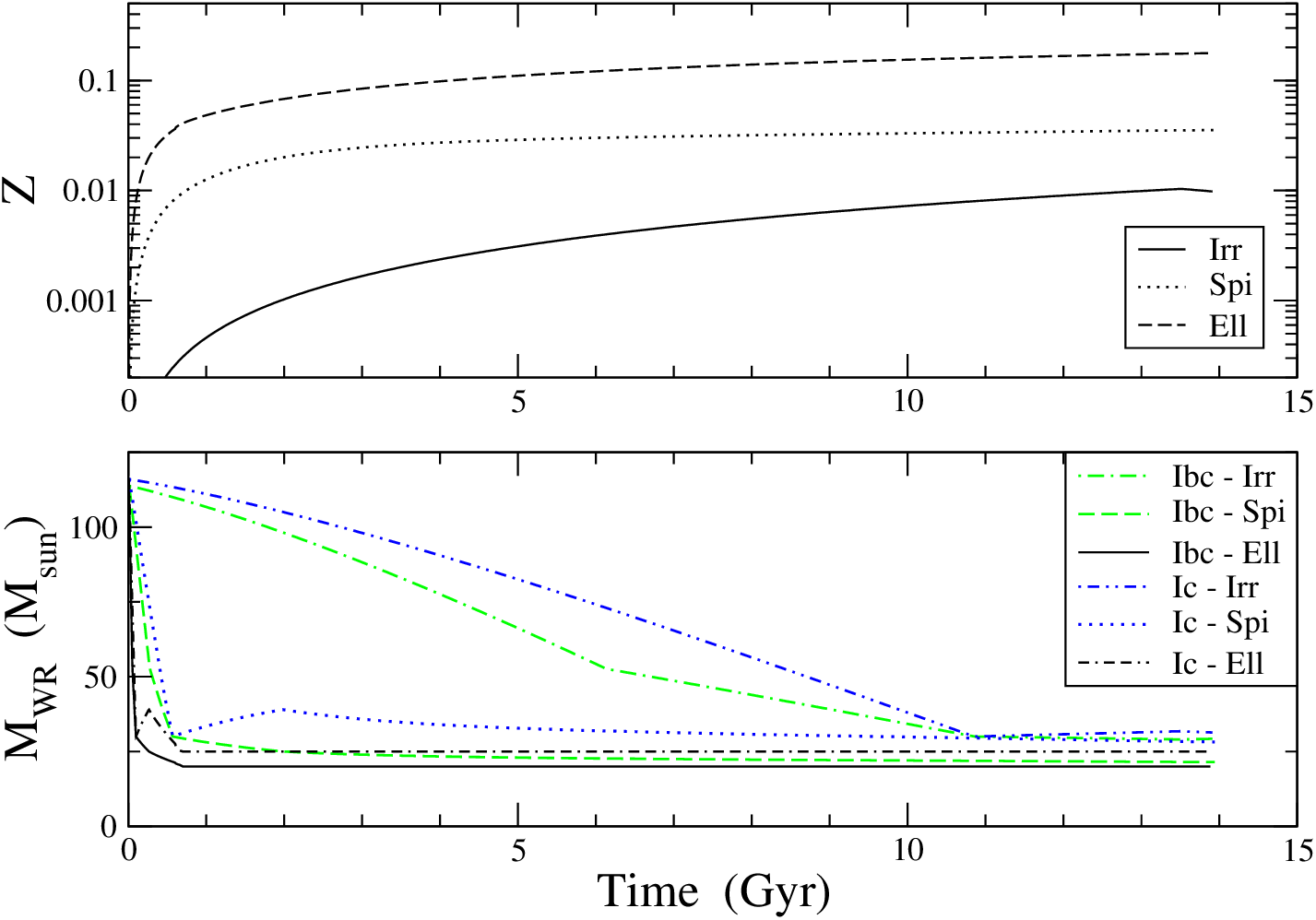} 
    \caption{Upper panel: evolution of the total metallicity as a function of time for the Model Irr  (solid line), 
             Model Sp (dotted line) and for a typical elliptical (dashed line).
             Lower panel: evolution of the minimal mass of WR progenitors  as a function of time for SNIbc 
             and SNIc in the Model Irr (green dotted-dashed line and blue double dotted-dashed  line), Model Sp 
             (green dashed line and blue dotted  line) 
             and for a typical elliptical (black solid line and black double dashed-dotted  line).} 
    \label{EvoZMwr}
    \end{center}
    \end{figure}

In Figure \ref{EvoZMwr} we show the evolution of the metallicity $Z$ for our galaxies and the corresponding evolution of the minimum WR mass, 
according to the relations of  Table 2 and 3. As one can see, owing to the milder increase of metallicity in the irregular galaxy, the $M_{WR}$ 
varies more gradually as a function of time in this galaxy than in the spiral one. On the other hand, the variation of $M_{WR}$ in an 
elliptical is very fast in the very early phases.

\begin{figure}
    \begin{center}
    \includegraphics[width=0.51\textwidth]{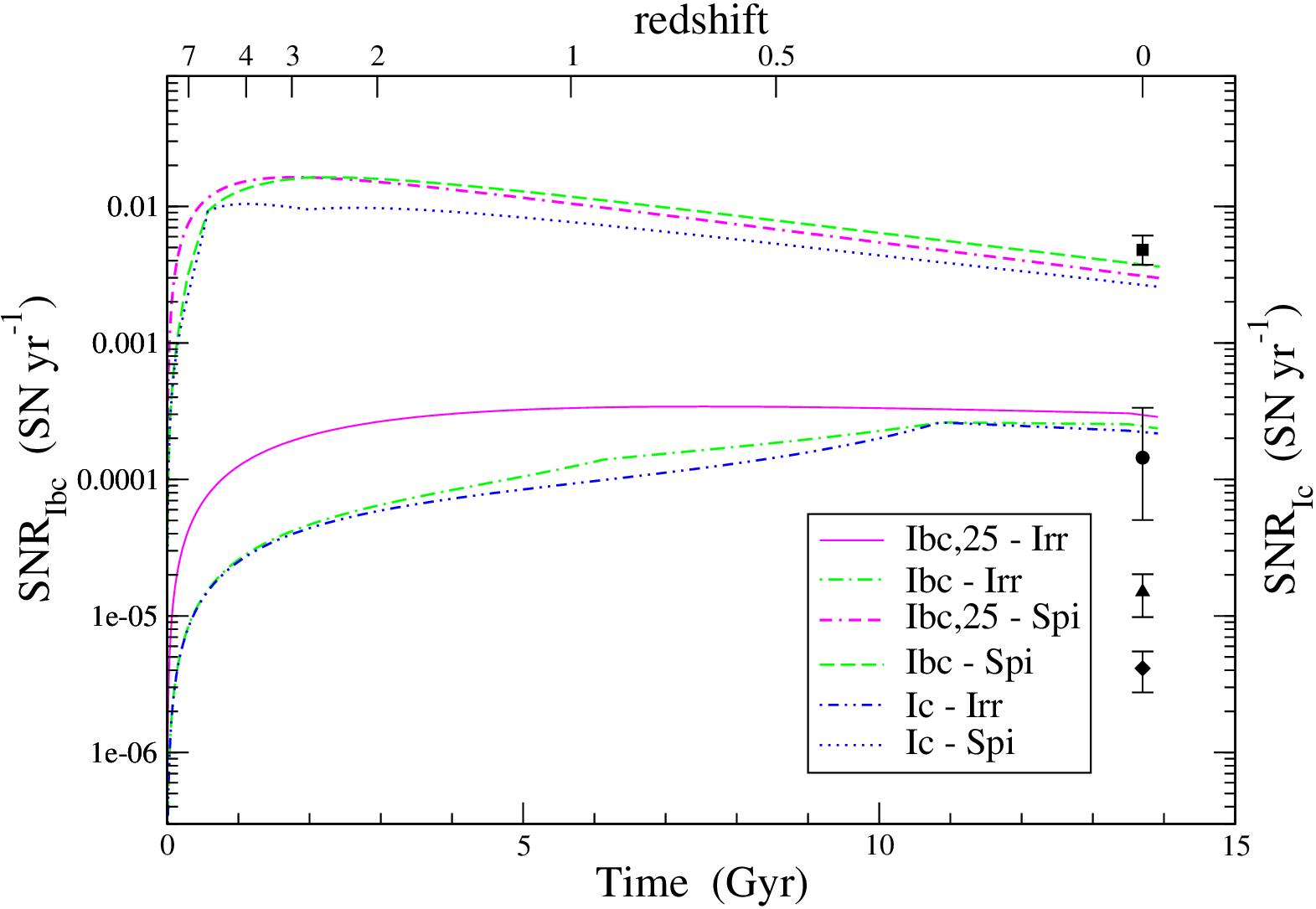}  
    \caption{Supernova Ib/c and Ic rates as a function of time and redshift for Model Irr (magenta solid line, green dotted-dashed  line 
    and blue double dotted-dashed  line) and Model Sp ( dotted-double dashed magenta line, green dashed line 
    and blue dotted line).
    The redshift of galaxy formation is $z_{f}=10 $ in a $\Lambda$CDM cosmology.
     The various model predictions for each rate depend on the different 
    assumptions concerning the SNIb/c and SNIc progenitors and their dependence on the initial stellar metallicity:
       $M_{WR}=25M_{\odot}$ (solid line for Model Irr and dotted-double dashed line for Model Sp), $M_{WR}=M(Z)$ 
       (dashed line for SNIbc and dotted line for SNIc in Model Sp; dashed- dotted line for SNIbc 
       and double dotted-dashed line for SNIc in Model Irr). 
       The points are the observed SN Ib/c rates, obtained by multiplying the observed rate per unit mass (Li et al. 2011) 
       by the present time stellar mass of the galaxy in Mod. Irr (circle) and Mod. Sp (square). 
       In the Figure is shown also the local GRB rate  provided by Guetta et al.(2005), triangle, 
       and Salvaterra et al. (2011), diamond.} 
    \label{snr}
    \end{center}
    \end{figure}

In Figure \ref{snr}, the SN Ib/c, Ic rates are shown as a function of time for Model Irr and Model Sp.
Different rates arising under different assumptions concerning the SNIb/c and SNIc progenitors and 
their dependence on the initial stellar metallicity (see sect. 2.1).
The percentage of SNe from binary systems represents $\sim$ 30\% of the total predicted rate.
The points in Figure \ref{snr} are the observed SN Ib/c rates for an irregular (circle) and  a spiral 
(square) galaxy provided by Li et al. (2011).

It is worth noting that the samples of SN-host galaxies usually include
(due to an observational bias) small numbers of irregulars and there is no
way to get out of this as long as the SN surveys will not include
a large number of such irregular galaxies. Following Li et al.(2011), one can
conservatively assume that the true value of the SN rate in Irregulars is
between Li et al. (2011) and Mannucci et al (2005) estimates.
In Figure \ref{snr} this uncertainty is included in the size of the error bars.

As one can see, the predicted present time rate of SNe Ib/care consistent, within the error bars, with the observed ones.

\begin{figure}
    \begin{center}
    \includegraphics[width=0.5\textwidth]{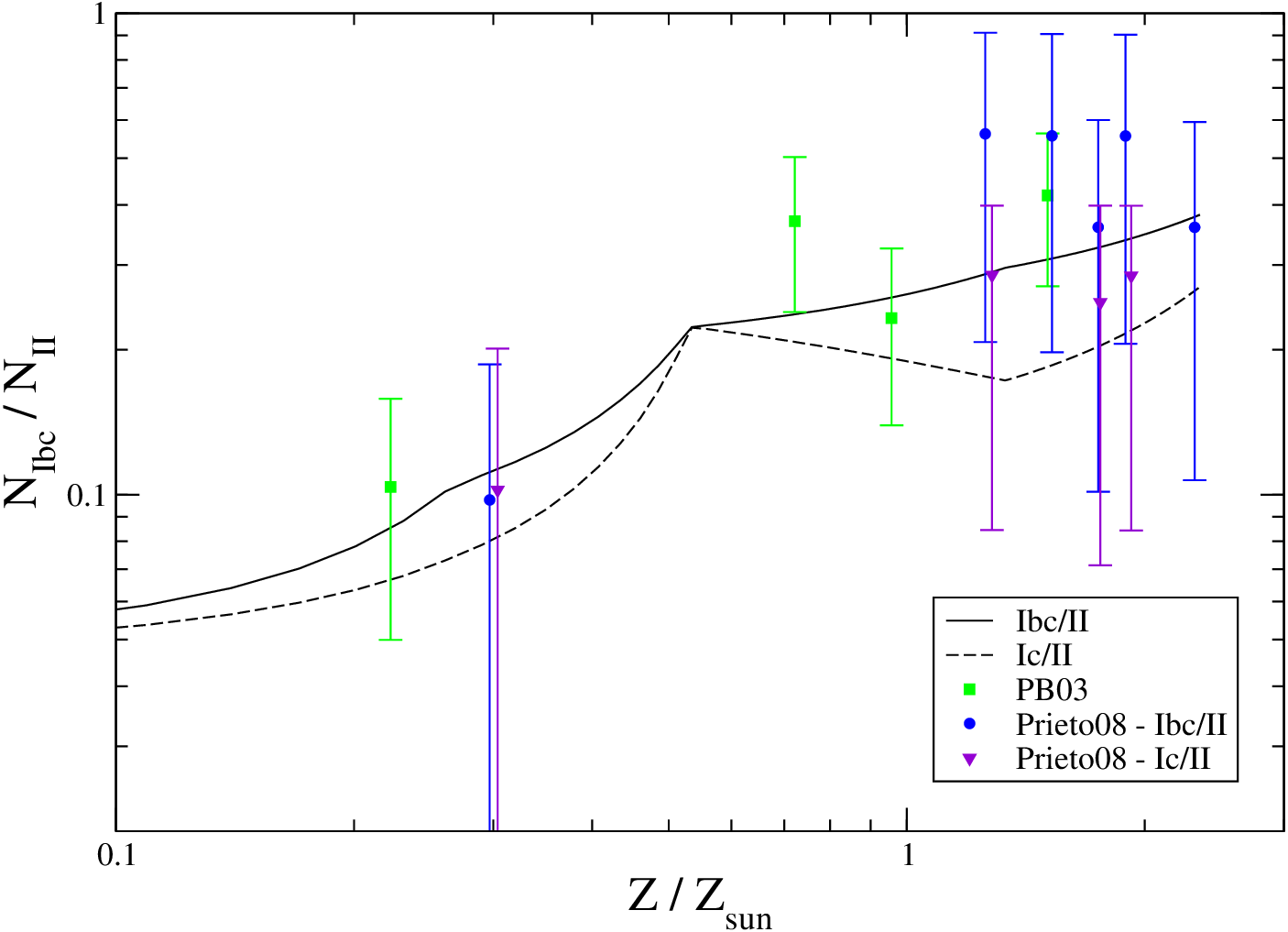}  
    \caption{Number ratio of SNIb/c and Ic to SNII as a function of metallicity of the host galaxy  
       (solid and dashed lines respectively). 
       The circles and the triangles down are the values obtained by Prieto et al. (2008) from directly 
       measured central metallicities for SNIbc and SNIc respectively while the
       squares are the results from Prantzos \& Boissier (2003) using absolute magnitudes as a proxy to host metallicities.} 
    \label{I/IIvsZ}
    \end{center}
    \end{figure}

 An interesting quantity, often shown in literature, is the ratio between 
the SN Ib/c and the SN II rate, 
$N_{Ibc}/N_{II}$ (i.e. Prantzos \& Boissier 2003; Boissier \& Prantzos 2009; Prieto et al. 2008; Smartt et al. 2009 and Smith et al. 2011).
In order to have a more consistent comparison with the data we have  plotted the $N_{Ibc}/N_{II}$ and $N_{Ic}/N_{II}$ ratios
as functions of  metallicity (see Figure \ref{I/IIvsZ}).
It is worth noting that when using $N_{Ibc}/N_{II}$ and $N_{Ic}/N_{II}$ vs metallicity, we cannot specify the galaxy type because 
the relation between mass and metallicity and the IMF are the same for all galaxy types.
As one can see from Figure \ref{I/IIvsZ}, our predictions are in good agreement with the data both in the case of SNe Ib/c (solid line) 
and SNe Ic (dashed line) and with previous calculations (Boissier \& Prantzos, 2009).

\section[]{The local $\mbox{GRB/SNI}_{bc}$ ratio} 

Figure \ref{snr} shows the existence of a good match between the \textquotedblleft expected\textquotedblright  SNIb/c 
\textquotedblleft local\textquotedblright rate, computed for spirals and irregulars, and the \textquotedblleft observed\textquotedblright SN 
rates derived by Li et al. (2011). This agreement assesses quantitatively the reliability of the prescriptions that have been used in 
Sections 2 and 3 to derive the SNIb/c rate as a function of time and provides more weight to the results that will be presented in Section 5.
When the metallicity effect on the minimum WR mass is taken into acount, there is a difference only for irregulars, mainly at early times, 
due to the slower growth of metallicity in these systems.

It is interesting to compute the ratio of Type Ib/c SNe to GRBs (in the local Universe) and to do that we should compare 
the theoretical SNIb/c rates, shown in Figure \ref{snr}, with the local rate of GRBs at the present time (triangle).  
The \textquotedblleft canonical\textquotedblright value for the latter quantity ranges between  
$\sim$ 0.5 GRB $\mbox{Gpc}^{-3}\, \mbox{yr}^{-1}$ (Schmidt 2001) to  $\sim$  1 event $\mbox{Gpc}^{-3} \mbox{yr}^{-1}$ 
(Guetta et al. 2005). 
In order to compare the SNIb/c and the GRB rate in the right units we use 
the local GRB rate  of 1.1 event $\mbox{Gpc}^{-3} \mbox{yr}^{-1}$ (Guetta et al. 2005) taking into account 
the local density of B luminosity, $\sim 1.2 \cdot 10^{8} \,\,L_{B,\odot}\,\, \mbox{Mpc}^{-3}$ (e.g. Madau, Della Valle \& Panagia 1998) 
and the B luminosity of the Milky Way, $2.3 \cdot 10^{10}\,\, L_{B,\odot}$. 
This approach gives $R_{GRB} \sim 2.1 \cdot 10^{-7} \,\,\mbox{yr}^{-1} $.
This \textquotedblleft observed\textquotedblright rate  has to be re-scaled 
by using the beaming factor, $f^{-1}_{b}$. 
The beaming factor accounts for the fact that a GRB does not light up the full 
celestial sphere but rather a fraction. 
There are several estimates of this parameter,  $f^{-1}_{b} \leq 10$ 
(Guetta \& Della Valle, 2007) for local and low luminosity GRBs, corresponding 
to $\theta > 25^{o}$, and 
$f^{-1}_{b} \sim 75-500$ (Guetta et al. 2005, Yonetoku et al. 2005; van Putten \& Regimbau 2003, Frail et al. 2001), for high-luminosity  GRBs, 
corresponding to beaming angles of $\sim  10^{-4}$.

If we conservatively assume,  for \textquotedblleft local\textquotedblright GRBs,  $f^{-1}_{b} \leq  75$, 
we derive a \textquotedblleft local\textquotedblright ratio GRB/SNIb/c of $\leq 3 \cdot 10^{-3}$ and 
$\leq 2 \cdot 10^{-2}$ in spirals and irregulars, respectively.
This is an expected result, since not all SNe Ib/c will end up as long GRBs. 

In Figure \ref{snr} we have shown the GRB rate provided by Guetta et al. (2005) in unit of $\mbox{yr}^{-1} $ and re-scaled using 
the beaming factor $f^{-1}_{b} = 75\pm 25$.

\section[]{The cosmic  $\mbox{SNI}_{bc}$ and GRB rates} 

\begin{figure}  
    \begin{center}
    \includegraphics[width=0.5\textwidth]{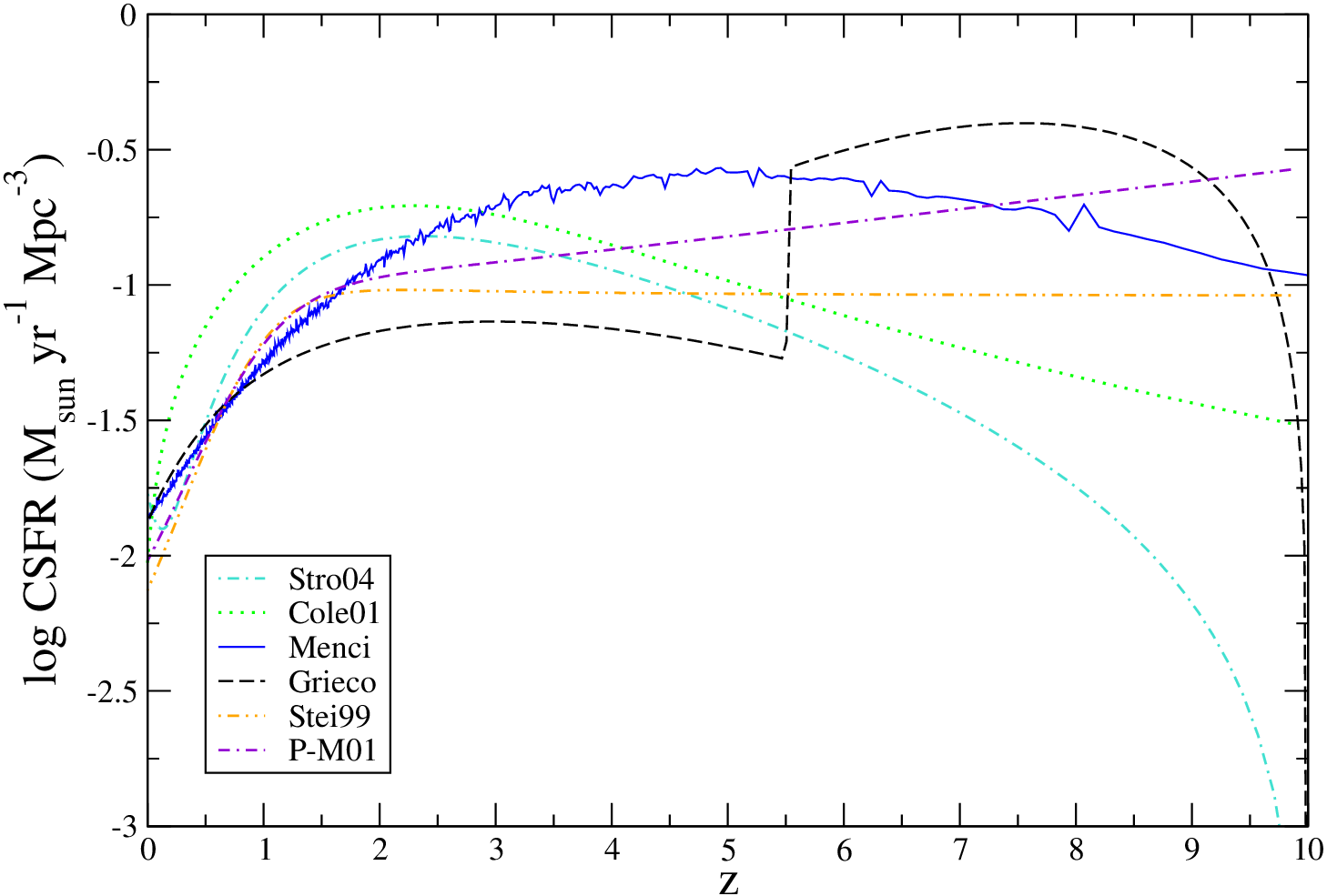} 
    \caption{Evolution of different cosmic star formation rates with redshift: 
             Menci, private communication (blue solid line), our model (black long-dashed line),
             Strolger 2004 (turquoise dashed-dotted line), Steidel 1999 
             (orange double dotted-dashed line), Porciani \& Madau 2001 (violet double dashed-dotted line). 
             The green dotted line is the fit (Cole et al. 2001) of the data collected by Hopkins (2004): this fit has then been extended up to redshift z=6.}
    \label{CSFR} 
    \end{center}
    \end{figure}

The cosmic SN and GRB rates are defined in an unitary comoving volume of the 
Universe. 
This definition is necessary to study the rates at high redshift 
where the morphology of the observed galaxies is not known. The cosmic rates refer, 
in fact, to a mixture of galaxies which can be different at every redshift. 
Both the cosmic Type Ib/c and GRB rates depend upon the SFR in galaxies 
but also on the galaxy and GRB luminosity functions. If these functions do not evolve with 
redshift then both the SN and GRB rates will trace the CSFR. 
On the contrary, the observed behaviour can be due to the evolution of the luminosity 
functions of galaxies (e.g. number density evolution). The CSFR has now been measured  up to very high redshift (z$\sim$8), 
especially thanks to galaxies hosting GRBs (see Kistler et al. 2009). 
 In Figure \ref{CSFR}, we show a revised version of the CSFR predicted 
by Calura \& Matteucci (2003) and obtained  by taking into account the evolution of galaxies of 
different morphological type (ellipticals, spirals and irregulars), as described in the previous sections. 
In particular, the CSFR has been computed by assuming a pure luminosity evolution of galaxies, in other words, 
the main parameters of the Schechter (1976) galaxy luminosity function have been kept constant with redshift.
To compute the CSFR we have adopted the following relation:
\begin{equation}
CSFR= \sum_{k}{\psi_{k}(t)\cdot n_{k}^{*}}\,\,\,\,\,\,\, [M_{\odot} \mbox{yr}^{-1} \mbox{Mpc}^{-3}],  
\end{equation}
where $k$ identifies a particular galaxy type (elliptical, spiral, irregular) and $\psi_{k}(t)$ represents the history of star 
formation in each galaxy, as shown in Figure \ref{figSFRmie}. 
The quantity  $n_{k}^{*}$ is the galaxy number density, expressed in units of $\mbox{Mpc}^{-3}$ 
for each morphological galaxy type and it has been assumed to be constant and equal to the present time one, as derived by Marzke et al. (1994).
This CSFR  is therefore obtained by assuming that all galaxies started forming stars at the same time and that there is no number density evolution; 
these are very simple assumptions but they can be useful to disentangle the effect of the SFR from that of the luminosity function and to compare  
these results with predictions from hierarchical galaxy formation models. This predicted CSFR shows a high 
peak of star formation at very high redshift due to the contribution of the ellipticals which have formed their stars very early. 
This predicted high redshift CSFR is probably too high and unrealistic since the number density of  
ellipticals at high redshift could have been overestimated. In other words, the hypothesis of no number density evolution could be incorrect. 
On the other hand, the predicted CSFR for redshift $z< 6$ seems underestimated relative to the data. This does not mean that our galactic SF 
histories are wrong but again it could be due to neglecting the number density 
evolution of galaxies. 
We have adopted also CSFRs computed in the framework of the
hierarchical clustering galaxy formation scenario, as well as the fit to the observed CSFR. 
In fact, in Figure \ref{CSFR} is also shown the CSFR obtained by Cole et al. (2001) 
 best fitting the data collected by Hopkins (2004) from 1995 onwards. 
This same parametric form has been used later on by many authors such as 
Hopkins \& Beacom (2006) and Blanc \& Greggio (2008), since it fits also more recent data up to redshift z=6.

It is worth noting that also all the other theoretical CSFRs shown in Figure \ref{CSFR} are underestimating the CSFR at intermediate and low redshifts.
To derive observationally the CSFR one should adopt some of the well known tracers of SF, in particular $H_{\alpha}$, $H_{\beta}$, UV continuum. 
In these wavebands the effect of dust cannot be neglected and therefore the dust correction is necessary to obtain the correct CSFR. 
The differences between corrected and uncorrected data are generally large, as shown by Strolger et al. (2004). 
In particular, uncorrected data tend to show a strong decline of the CSFR for  $z >2$, whereas the corrected data show an almost constant 
CSFR for $z>3$. Another important effect in the derivation of the CSFR is related to the uncertainty in the
faint end of the luminosity function of galaxies.

\begin{figure}
    \begin{center}
    \includegraphics[width=0.5\textwidth]{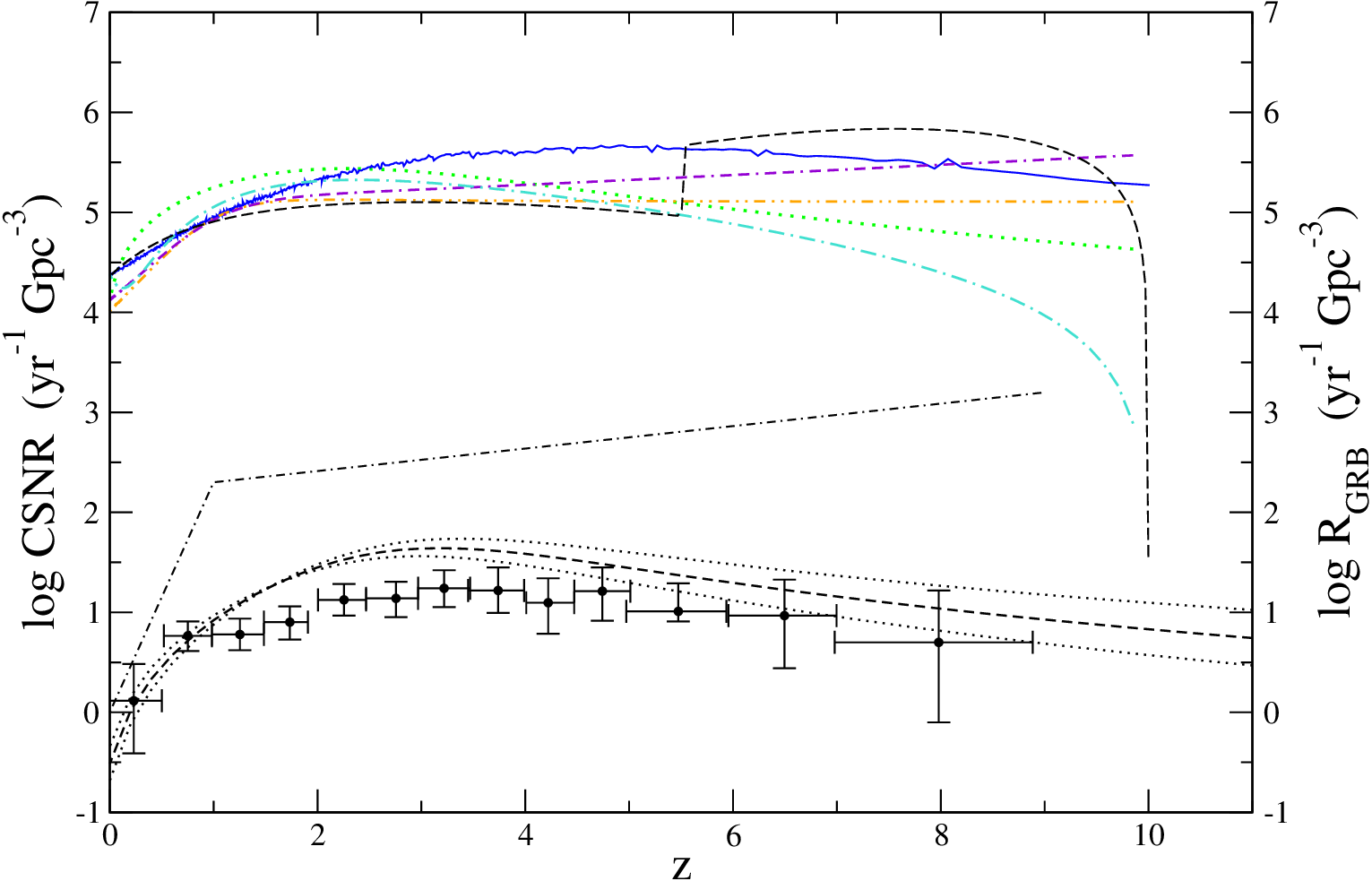} 
    \caption{Comparison between the cosmic predicted Type Ib/c SN rates computed by means of all the CSFRs of Figure 6 and 
             the number of observed GRBs at different redshift provided by Wanderman \& Piran (2010), {\itshape Swift} data, 
             (black circles with error bars) and Matsubayashi et al. (2005) (black dashed-dotted line in the lower part of the Figure). 
             The short-dashed and the double dotted black line, below the Matsubayashi et al. (2005) rate, 
             represent the best fit and the upper and lower limit, respectively, of the cosmic GRB rate obtained by Salvaterra et al. (2011). 
             The CSNRs Ib/c are computed by means of CSFRs shown in Figure 6 and are indicated with the same symbols.}
    \label{CSNR_GRB} 
    \end{center}
    \end{figure}

\begin{figure}
    \begin{center}
    \includegraphics[width=0.5\textwidth]{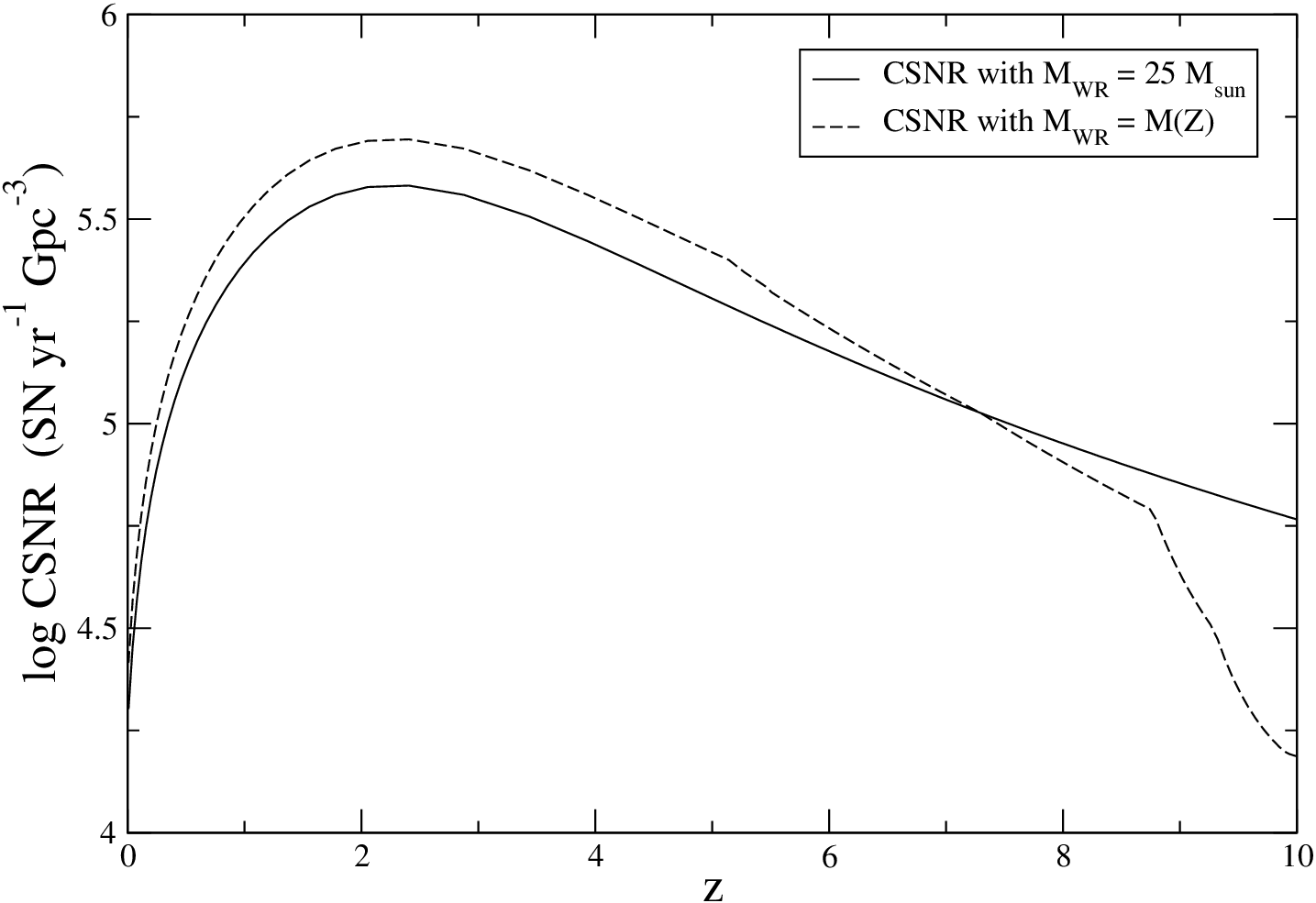} 
    \caption{Comparison between the cosmic predicted Type Ib/c SN rates computed by means of the CSFR of Cole et al. (2001) 
             and the different assumptions on $M_{WR}$: 
             solid line refers to a constant $M_{WR}$ whereas dotted line refers to the case of  $M_{WR}=M(Z)$.}
    \label{CSNR_Z} 
    \end{center}
    \end{figure}

 By means of these different CSFRs we have then computed the cosmic SN Ib/c rate
shown in Figures \ref{CSNR_GRB} 
where it is reported also the observed cosmic GRB rate. The adopted progenitors for SNe Ib/c are assumed to be single WR stars with constant 
minimum mass of $25M_{\odot}$ plus binary systems, as described in section 2.1. 
 As one can see, the theoretical error in the CSNR increases towards high and very high redshift and it is roughly a factor of ten at z=6. 
Clearly at these high redshifts ($z>6$) the uncertainties are still too large too draw any conclusion.
In Figure \ref{CSNR_Z} we show the predicted cosmic Type Ib/c SN rate obtained by adopting the Cole et al. (2001) CSFR and 
$M_{WR}$ depending on Z, all the other assumptions being the same. 
Here we have considered the cosmic evolution of Z, and in particular we assumed the Z vs. time 
evolution typical of an elliptical galaxy of $10^{11} M_{\odot}$ as shown in Figure \ref{EvoZMwr}; 
this is because by weighting the Z vs. time of each galaxy on their number density, 
the Z vs.time relation of the ellipticals dominates at all redshifts. 
In fact, spheroids are very likely to be the responsible for the production of 
the bulk of metals in the Universe (see Calura \& Matteucci 2004).
As one can see in Figure \ref{CSNR_Z}, the effect of metallicity on the Type Ib/c SN progenitors is stronger at early times and produces a lower cosmic 
Type Ib/c SN rate.
This effect would be similar if applied to all the CSFRs of Figure \ref{CSFR} 

Coming back to Figure \ref{CSNR_GRB}, a visual  inspection of this figure confirms that GRBs originating from the explosion of massive stars are only a tiny 
fraction of SNe Ib/c class. In particular, the comparison of the SNIb/c rates with the Matsubayashi et al. (2005) semi-empirical track, suggests 
the ratio GRB/SNIb/c to be $\sim  10^{-4}$ in the local Universe  and to increase up to $\sim  10^{-3}-10^{-2}$ all over the redshift range z= 1 $\div$ 8.  
Interestingly enough, the GRB/SNeIb/c ratio at z $\sim$ 0 nicely reproduces the \textquotedblleft observed\textquotedblright ratio between the local GRB rate and the SNIb/c rates 
obtained by other authors:  $\sim 1 GRB\,\, \mbox{Gpc}^{-3}\,\, \mbox{yr}^{-1}$ (Guetta et al. 2005) and 
$ \sim 2 \cdot 10^{4} SNIb/c\,\, \mbox{Gpc}^{-3}\,\, \mbox{yr}^{-1}$ 
(Guetta \& Della Valle 2007), respectively.  After taking these figures at their face values, we conclude that \textquotedblleft local\textquotedblright 
and \textquotedblleft low luminosity\textquotedblright GRBs 
($L \le 10^{49}\, ergs^{-1}$) barely need the correction for beaming and therefore we can infer that they emit almost isotropically. This result 
is in good agreement  with observations (admittedly on scanty statistic).  For example, for GRB 060218 Soderberg et al. (2006) find $\theta > 70 deg$, 
for GRB 031203, $\theta > 30 deg$ (Malesani, private communication) and $\theta > 25 deg$  (corresponding to $f^{-1}_b \le 10$) for  
\textquotedblleft low luminosity\textquotedblright GRBs population (Guetta \& Della Valle, 2007).
The increasing ratio GRB/SNIb/c $\sim 10^{-3}-10^{-2}$ for $z \sim 1$,  in the case of the Matsubayashi et al. (2005) GRB rate, may suggest either a different behavior 
for \textquotedblleft cosmological\textquotedblright 
GRBs due to the existence  of a different GRB population (e.g. Bromberg et al. 2011 and references therein) characterized by a larger beaming factor, 
likely of the order of  $f^{-1}_b \sim20-200$ (corresponding to a jet opening angle of $\sim$ 20deg-6deg) or the Matsubayashi et al. semi-empirical 
track is still affected by the obvious bias which favors the discovery at high-$z$ of only highly beamed GRBs.
Similar conclusions (short of a constant) can be obtained by comparing the SNIb/c rate track with the GRB track by Wanderman \& Piran (2010) and with 
that by Salvaterra et al. (2011).
In particular,  the Salvaterra et al. (2011) rate is derived from a redshift complete sample of bright {\it Swift} 
GRBs under the assumption that GRBs did not experience luminosity evolution with redshift.  
It is worth noting that on the basis of the current studies, it is not possible to distinguish between a pure density and a pure luminosity evolution.
In general, there is no agreement among different authors on this issue.
Butler et al. (2010) suggest that pure density evolution models produce the observed number of GRBs at high-z,
but in other works the luminosity evolution is used to explain the GRB rate 
(i.e. see Salvaterra et al. 2009b, Petrosian et al. 2009) increasing  faster than some CSFR, such as that of   Hopkins \& Beacom (2006) .
This is because the Hopkins \& Beacom (2006) CSFR is decreasing at high z. However, this behaviour of the CSFR needs to be confirmed by 
more data and at the moment we cannot exclude the CSFR to be flat at high z.

\section[]{Conclusions}

 In this paper we have computed the Type Ib/c SN rates expected at the present time in irregular and spiral galaxies of different 
masses with the aim  of predicting the variation with redshift of the SNIb/c rate based on successful models for 
the chemical evolution of irregulars and spirals.
We considered both single WR stars and massive stars in binary systems as SN Ib/c progenitors. 
We used stellar evolution results indicating that the minimum mass of WR stars is a function of the stellar metallicity thus suggesting 
a higher rate of SNe Ib/c in more metal rich galaxies. 
Then we considered various CSFRs as 
functions of cosmic time, both theoretically and observationally derived, and computed the cosmic Type Ib/c SN rates expected from the 
assumptions on Type Ib/c 
SN progenitors. These cosmic Type Ib/c SN rates were then compared to the observationally derived cosmic GRB rate.
Our main conclusions can be summarized as follows:

\begin{itemize}

\item by taking into account WR progenitors depending on the metallicity and a fraction of massive close binary systems equal to 15\% of all 
massive stars as SN Ib/c progenitors, it is possible to reproduce the present observed Type Ib/c SN rate both in dwarf metal poor irregular  and in 
spiral galaxies. 
It is worth noting that the galactic evolution models adopted here are well reproducing the main chemical properties of these galaxies. 

\item If a dependence on stellar metallicity is assumed for the WR stars, differences arise in the Type Ib/c SN rates only at early evolutionary 
times in galaxies. Negligible differences are produced on the predicted local rates.

\item We have compared the local observed Type Ib/c rates in spirals and irregulars with the local GRB rate and derived a local  ratio GRB/SNe Ibc of 
$\sim 3 \cdot 10^{-3}$. As expected, only a fraction of these SNe gives rise to GRBs.
 
\item We took various CSFR histories and
computed the cosmic Type Ib/c SN rates.
Also in this case we considered both a constant minimum WR mass and a mass varying with metallicity. The effect of the dependence of $M_{WR}$ 
on the metallicity is to predict lower cosmic Type Ib/c SN rates at very high redshift.
 We have then compared the cosmic Type Ib/c SN rates with the cosmic GRB rate
derived from {\itshape Swift} data
and found that the  ratio GRB/SNe Ibc $\sim 10^{-4}$. 
This confirms previous results that only a small fraction of all SNe Ib/c 
gives rise to GRBs, but our factor is smaller than what found in 
Bissaldi et al. (2007). The reason for this resides in the fact that we 
have adopted the recent GRB cosmic rate derived from Swift data, whereas
in Bissaldi et al. (2007) the cosmic GRB rate was derived on 
the basis of semi-empirical estimates (Matsubayashi et al. 2005). On the other hand, if we compare our cosmic SN rates with the Matsubayashi et al. (2005)
 rate we confirm the results of Bissaldi et al. (2007), indicating a ratio GRB/SNIb/c rates of $\sim 10^{-3} - 10^{-2}$.

\item Studies of GRBs and their hosts have revealed to be extremely important to trace galaxy evolution at very high redshifts, 
although the interpretation of cosmic diagrams is difficult since it involves assumptions on the luminosity function of both galaxies and GRBs. 
It is interesting to note that Salvaterra \& Chincarini (2007) pointed out that by adopting the CSFR derived by Cole et al. (2001) and assuming  
a GRB luminosity function independent of redshift, one largely underestimates the number  of high redshift GRBs detected by Swift. 
This fact could be interpreted in two ways: either the characteristic luminosity of GRBs increases with redshift or the CSFR at very high redshift is higher 
than in Cole et al. (2001). We have shown that a high CSFR can be achieved by means of monolithic like models of ellipticals producing stars 
at a very high rate and at very high redshift. 
However, no firm conclusions can be drawn on the CSFR at very high redshift because of the large uncertainties due to dust corrections.

\end{itemize}

\section*{Acknowledgments}
V.G.  thanks Anah\'i Granada, Cyril Georgy and Luca Vincoletto for many useful discussions.

\end{document}